**Using multicomponent recycled electronic waste alloys to produce high entropy alloys**

José M. Torralba [1,2], Diego Iriarte [2], Damien Tourret [2], Alberto Meza [2]

1) Universidad Carlos III de Madrid
2) IMDEA Materials Institute

**ABSTRACT**

The amount of electronic waste (e-waste) recycled worldwide is less than 20% of the total amount produced. In a world where the need for critical and strategic metals is increasing almost exponentially, it is unacceptable that tons of these elements remain unrecycled. One of the causes of this low level of recycling is that recycling is based on an expensive and complex selective sorting of metals. Extracting all metals simultaneously is much simpler and if this were done, it would significantly increase the recycling rate. Meanwhile, it was demonstrated that high entropy alloys (HEAs), which are in great demand in applications where very high performance is required, can be made from mixtures of complex alloys, hence reducing their dependence on pure critical metals. Here, we show that it is possible to obtain competitive HEAs from complex alloy mixtures corresponding to typical electronic waste compositions, combining two needs of high interest in our society, namely: to increase the level of recycling of electronic waste and the possibility of developing high-performance HEAs without the need of using critical and/or strategic metals. To validate our hypothesis that e-waste can be used to produce competitive HEAs, we propose an alloy design strategy combining computational thermodynamics (CalPhaD) exploration of phase diagrams and phenomenological criteria for HEA design based on thermodynamic and structural parameters. A shortlist of selected compositions are then fabricated by arc melting ensuring compositional homogeneity of such complex alloys and, finally, characterised microstructurally, using electron microscopy and diffraction analysis, and mechanically, using hardness testing.

## 1. Introduction

Nineteen years have passed since Cantor et al. [1] and Yeh et. al. [2] introduced the concept of high entropy alloys (HEAs) and multi-principal element alloys. Today, HEAs have become a promising field of knowledge, with more than 10 000 scientific papers published according to most extended databases (Scopus, SCI). If we include the mentions to "multicomponent alloys" we can find more than 200 000 references (SCI). Among these papers, interesting reviews, such as the one by Miracle and Senkov[3] provide a complete state-of-the-art on the subject. Other general and critical reviews on the topic can be found in[4–8]. HEAs have demonstrated their ability to cover a wide range of properties. In[9], applications related to electrical or thermal properties are highlighted, while in[10–12] applications in extreme corrosion conditions are discussed. It is also clear that HEAs are promising materials due to their good mechanical performance [13,14] including at high[3] or cryogenic temperatures[15]. Other reviews related to mechanical behaviour deal with their deformation behaviour[16] or fracture resistance[17]. Two books were also written on the subject[18,19]. In 2021, a report (available in open access)[20] was published by The Minerals, Metals & Materials Society (TMS), written by 15 international experts and funded by two US defence agencies; this report defined the pathways for harnessing the revolutionary potential of HEAs. Several reviews [21,22,23] have also addressed powder metallurgy of HEAs, one of them focusing on additive manufacturing[23].

An important aspect that hinders the use of HEAs at industrial scale is their typical high amount in critical metals, which are defined as metals that combine a possible lack of availability with an extremely high price according to the classification of the European Union[24]. Therefore, the development of competitive HEAs that avoid exploiting new mining sites and importing critical raw materials, e.g. critical metals (as direct feedstock), is a major high-impact challenge. Overcoming this challenge could extend the use of HEAs to many industries, and greatly reduce the negative impact of raw material availability and cost.

Additionally, our society is facing another problem of enormous importance, also related to the challenges of sustainable development, which is the exponential growth of electronic waste (e-waste) with very limited recyclability. According to[25], in 2019, 53.6 Mt of e-waste was generated and the forecast for 2030 is 74.7 Mt. From those, only 9.3 Mt (17.4%) will be recycled. Between 30 and 32% in weight of the recycled e-waste are metals[26]. This low recycling rate is directly related to the difficulty of selective recycling. The recycling of metals from e-waste faces problems linked to the extraction of each element separately, forcing the use of different selective leaching agents, costly separation and purification processes or high-temperature pyrometallurgical processes. Yet, for the purpose of using these recycled metals as raw materials to produce HEAs, selective extraction is not necessary. Therefore, the recycling possibility of e-waste, considering all the elements together as a complex alloy, could tremendously simplify and facilitate the recycling of electronic products[27]. These complex "multi-component recycled e-waste alloys", made up of many metals, could be used to produce HEAs using similar alloy design concepts as those proposed in[28] by mixing "commodity" alloy powders.

Torralba and Kumaran[28] recently showed that by properly combining different base alloys (as different sources of alloying elements), competitive high entropy alloys could be obtained. In[29] this hypothesis was also demonstrated using spark plasma sintering to produce the bulk material. Following up from that study, here we assert that both recycled scrap and commodity alloys could be similarly used as raw materials to design and produce competitive HEAs.

Previous studies[26,30,39–41,31–38] have established typical compositions for multi-component recycled e-waste alloys. Here, we study the feasibility of obtaining high entropy alloys by mixing some of these typical compositions for e-waste coming from the disposal of smartphones, laptops, Central processing unit (CPUs) and Li-ion batteries, possibly mixing them too with commodity alloys (CoCr75 and 316L).

Our hypothesis is that HEAs with potentially exceptional properties can be made from the mixture of complex alloys extracted from e-waste. Proving this hypothesis would open up the possibility of using HEAs extensively in industries where alloys with outstanding properties are required, without the need of importing critical and/or strategic metals. On the other hand, it would also open up a promising new route for using electronic waste through cheaper and easier recycling (mined as a whole) than is currently done (selective recovery of metals). Methods

From various composition proposals [26,30,39–41,31–38] for multicomponent e-waste alloys found in the literature (see Table 1), four alloys were identified that can be considered as Co/Ni-based complex alloys (Table 2). We followed three criteria to choose the alloys representing the e-waste starting materials (Table 1): 1) they had to come from very abundant electronic waste (smart phones, laptops, ion-lithium batteries, and CPUs). These sets of compositions also allow to play with three large families of compositions, based on Cu, Ni and Fe. The presence of Al in some of these alloys is interesting for the formation of eutectic high entropy alloys. 2) We have considered the use of some of these alloys, once the Cu and/or Al has been extracted by selective leaching, since there are currently recycling companies that are only interested in these metals, leaving residues with the rest of the components. 3) In order to complete the possible final alloys, we have relied on the use of scrap compositions from two widely used "commodity" compositions such as 316L stainless steel and CoCr75 alloy. The four target alloys (Table 2) are intended to play with transition elements, typical in many HEAs, together with Al which allows the possibility of eutectic HEAs. For the purpose of this work to demonstrate the feasibility of the proposed hypothesis, we have reproduced the projected compositions from the literature for the e-waste alloys.

Carrying out some experiments with combinations of the proposed alloys (Table 1) and using the HEAPS program (a tool for the design and study of HEAs based on semi-empirical parameters)[42], we identified four possible combinations, which, according to some of the threshold criteria proposed by HEAPS, allow to obtain HEAs with a full solid solution of and FCC phase or a combination of FCC and BCC phases (see Tables 2 and 3). Table 3 summarizes some of the possible characteristics of the selected e-waste-based HEAs according to different study parameters used by reference [42].

The four alloys were produced by vacuum arc melting (VAR) under a high purity Ar atmosphere. Four re-melting processes were carried out to ensure microstructural homogeneity. A total of three Ar purges were performed prior to each of the melting processes. Table 4 summarizes the final compositions measured in the four alloys by compositional analysis using Energy Dispersive Spectroscopy (EDS in FEG-SEM, Apreo 2S LoVac). As observed, changes in composition have taken place due to sublimation. In particular, final alloys are (almost) completely depleted in Mn compared to the original target compositions. Table 4 also lists the phase prediction for the original pre-selected parameters[42], which is essentially similar to that with the final compositions. In support of the HEAPS-based alloy selection, we also performed thermodynamic equilibrium (lever rule and phase diagram) calculations using the CalPhaD method (software: ThermoCalc, databases: TCHEA6 and TCNI8), using final compositions (Table 4).

A Field Emission Gun Scanning Electron Microscope (FEG-SEM, Apreo 2S LoVac) equipped with energy dispersive X-ray spectroscopy (EDS) and Nordlys Electron Backscatter Diffraction (EBSD) detectors was also used to characterise structural and microstructural features. The samples were mechanically ground and polished with a series of diamond pastes down to 1 μm grit, followed by a final polishing with an oxide particle suspension (OPS, 0.04 μm) to improve the sample surface for EBSD analysis. The samples were also tested for hardness using a Vickers macro-hardness tester (using an INNOVATEST instrument) with a load of 1 kg.

**Table 1**. Possible complex e-waste alloys and used commodity alloys (% at.).

| Devices | Cu | Al | Fe | Sn | Ni | Co | Si | Mn | Cr | Mo | Ref. | Alloy |
|---|---|---|---|---|---|---|---|---|---|---|---|---|
| Smart phones | 28.50 | 49.10 | 12.65 | - | - | 9.74 | - | - | - | - | 33 | **A** |
| Laptops | 29.48 | 49.51 | 16.77 | 0.38 | - | 3.78 | - | - | - | - | 33 | **B** |
| Li-ion batteries | 4.83 | 26.55 | - | - | 48.82 | 9.55 | - | 10.25 | - | - | 38 | **C** |
| Smart phones (*) | - | 68.68 | 17.69 | - | - | 13.62 | - | - | - | - | 33 | **D** |
| Laptops (*) | - | 70.29 | 23.81 | 0.54 | - | 5.37 | - | - | - | - | 33 | **E** |
| Li-ion batteries (*) | - | 27.90 | - | - | 51.30 | 10.03 | - | - | 10.77 | - | 38 | **F** |
| Smart phones (**) | - | - | 56.50 | - | - | 43.50 | - | - | - | - | 33 | **G** |
| CPUs (**) | - | - | 32.56 | 2.70 | - | - | 64.74 | - | - | - | 31 | **H** |
| Li-ion batteries (**) | - | - | - | - | 71.15 | 13.92 | - | - | 14.94 | - | 38 | **I** |
| **Commodities** | **Cu** | **Al** | **Fe** | **Sn** | **Ni** | **Co** | **Si** | **Mn** | **Cr** | **Mo** | | |
| 316L | - | - | 68 | - | 12 | - | - | - | 19 | 1 | - | **J** |
| CoCrF75 | - | - | 1 | - | - | 62 | - | - | 33 | 4 | - | **K** |

(*) In these alloys Cu has been removed from the original e-waste composition.
(**) In these alloys Cu and Al has been removed from the original e-waste composition

**Table 2.** E-waste alloys obtained from mixes of alloys from Table 1.

| | | At. (%) | | | | | | | | | |
|---|---|---|---|---|---|---|---|---|---|---|---|
| **E-waste alloys** | **Mixture (at. %)** | **Cu** | **Al** | **Fe** | **Sn** | **Ni** | **Co** | **Si** | **Mn** | **Cr** | **Mo** |
| **E-waste 1** | 10**A**+10**B**+80**C** | 9.7 | 31.1 | 2.9 | - | 39.1 | 9.0 | - | 8.2 | - | - |
| **E-waste 2** | 5**A**+5**B**+90**C** | 7.2 | 28.8 | 1.5 | - | 44.0 | 9.3 | - | 9.2 | - | - |
| **E-waste 3** | 6**D**+24**E**+30**F**+30**J**+10**K** | - | 29.4 | 27.2 | 0.1 | 19.0 | 11.4 | - | 3.2 | 8.9 | 0.8 |
| **E-waste 4** | 60**G**+20**H**+20**I** | - | - | 40.4 | 0.5 | 14.3 | 28.9 | 12.9 | 3.0 | - | - |

**Table 3**. E-waste alloys and HEA features predictions[42].

| Semi-empirical parameters (according to [42]) | $\Delta H^m$ (kJ·mol$^{-1}$) - $\delta r$ (%) [43-44] | $\Omega$ (-) - $\delta r$(%) [44] | $\gamma$ (-) [6] | VEC (-) [45] | $\Delta\chi^P$ (-) [46] (*) | PSFE (at. %) - VEC (-) [47] | $\delta\chi^A$ (%) – $\delta r$ (%) [48] |
|---|---|---|---|---|---|---|---|
| **Established criteria** | SS 0.5 < $\delta r$ < 6.5 and −17.5< $\Delta H^m$ <5<br><br>IM/BMG $\delta r$ > 6.5 or −17.5 > $\Delta H^m$ or $\Delta H^m$ > 5 | SS 1.1 ≤ $\Omega$ and $\delta r$ ≤ 6.6<br><br>IM $\Omega$ < 1.1 or $\delta r$ > 6.6 | SS $\gamma$ ≤ 1.175<br><br>IM/BMG 1.175 < $\gamma$ | BCC VEC < 6.87<br><br>FCC 8 ≤ VEC<br><br>FCC +BCC 6.87 < VEC < 8 | TCP Phase 0.133 < $\Delta\chi^P$<br><br>TCP Free $\Delta\chi^P$ < 0.117 | Sigma Phase 6.88≤VEC≤7.84 and PSFE > 42.5<br><br>Sigma Free PSFE < 22.5 or VEC < 6.88 or VEC > 7.84 | Laves Phase 7 < $\delta\chi^A$ and 5 < $\delta r$<br><br>Laves Free $\delta\chi^A$ <7 or $\delta r$ > 5 |
| **E-waste 1** | SS | SS | SS | BCC+FCC | TCP Phase | Sigma Free | Laves Free |
| **E-waste 2** | SS | IM | SS | BCC+FCC | TCP Phase | Sigma Free | Laves Free |
| **E-waste 3** | SS | SS | IM/BMG | BCC | Uncertain | Sigma Free | Laves Free |
| **E-waste 4** | IM/BMG | SS | IM/BMG | FCC | TCP Phase | Sigma Free | Laves Free |

SS: solid solution; IM: intermetallic; BMG: bulk metallic glasses. VEC: valence electron concentration number; PSFE: paired sigma-forming element parameter. (*) Note: the $\Delta\chi^P$ parameter tends to be inaccurate for high Al content, as in alloys 1-3, according to [42]

**Table 4.** E-waste alloys actual measured compositions after arc melting process.

| E-waste alloys | Phase prediction | At. (%) | | | | | | | | | |
|---|---|---|---|---|---|---|---|---|---|---|---|
| | | Cu | Al | Fe | Sn | Ni | Co | Si | Mn | Cr | Mo |
| **E-waste 1** | BCC+FCC | 9.0 | 33.4 | 3.6 | - | 43.7 | 10.3 | - | - | - | - |
| **E-waste 2** | BCC+FCC | 8.6 | 29.2 | 1.9 | - | 49.5 | 10.8 | - | - | - | - |
| **E-waste 3** | BCC | - | 29.1 | 28.8 | 0.2 | 19.3 | 11.8 | - | 0.6 | 9.5 | 0.7 |
| **E-waste 4** | FCC | - | - | 41.3 | 0.6 | 14.3 | 30.9 | 12.9 | - | - | - |

## 2. Results

### 2.1. Thermodynamic calculations

Figure 1 shows the CalPhaD-calculated lever rule (namely: equilibrium volume fraction of phases vs temperature) for the 4 alloys, using the TCHEA6 database, considering the actual compositions after arc melting (Table 4).

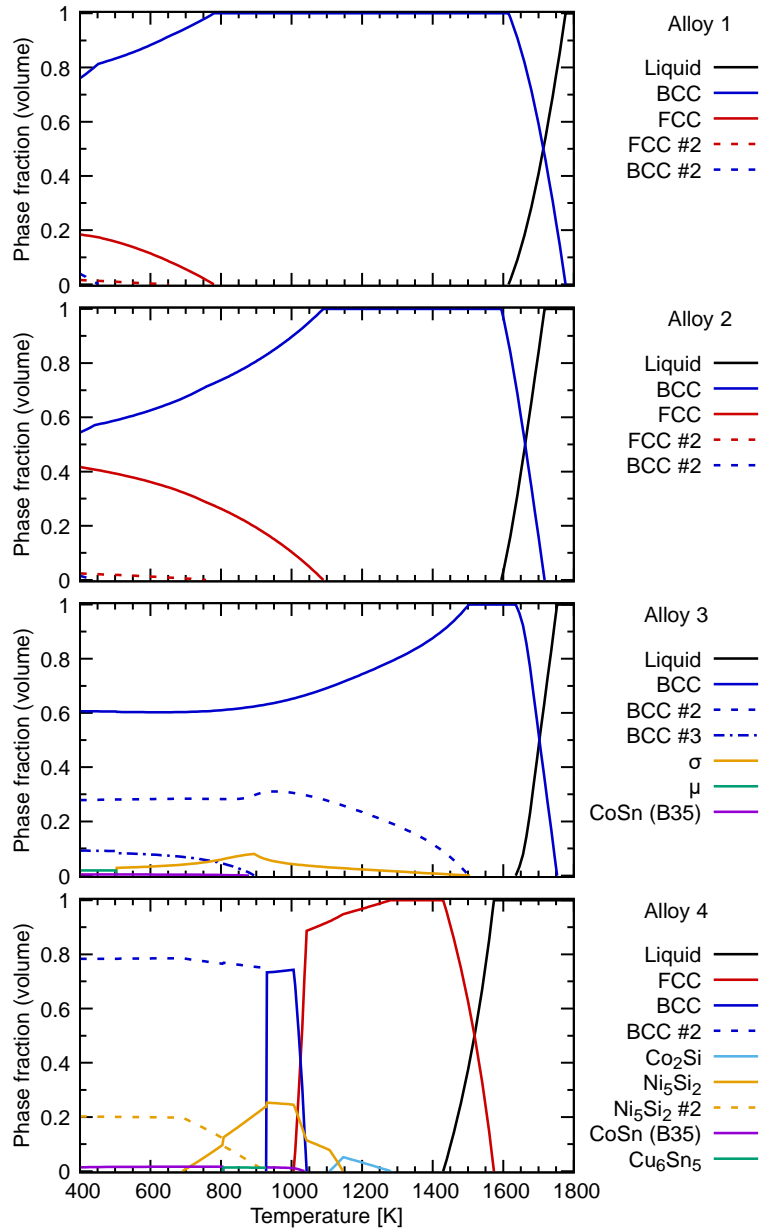

Figure 1. CalPhaD-calculated lever rule (phase fraction at thermodynamic equilibrium vs temperature) for the four e-waste alloys (Table 4).

### 2.2. Structural and microstructural analysis.

Figure 2 shows the XRD patterns obtained for all the e-waste alloys studied. We identified a simple BCC phase for two of the alloys (e-waste 1 and 3), a biphasic material for e-waste alloy 2 (FCC and HCP, where FCC is a minority phase) and also a biphasic material for e-waste 4 (FCC and BCC).

Figures 3 to 6 show the inverse pole figures (IPF-z) for the four developed e-waste alloys and the phase maps obtained by EBSD. E-waste alloys 1 and 3 exhibit a monophasic alloy (BCC) microstructure. E-waste alloys 2 and 4 exhibit a dual phase structure: FCC + HCP for e-waste 2, and FCC + BCC for e-waste 4. Analysing the lamellar area of alloy 2 at higher magnification (Figure 7), the unknown area (not indexed at lower magnification in Figure 4) can be identified as HCP. This is in good agreement with the XRD results. Figures C5 to C9 (in the supplementary data) show the complete IPF images with the phase map for the four e-waste alloys, also including a higher magnification of e-waste alloy 2.

Figures C1 to C4 (in the supplementary data) show additional SEM images of the four e-waste alloys produced by arc melting and a mapping distribution of the different alloying elements. e-waste alloys 1 and 4 have a dendritic structure; in e-waste 1, Fe, Cu and Al are most segregated in the dendrites, while Co and Ni are well distributed throughout the microstructure; and in e-waste 4, Fe and Co are segregated in the dendrites, while Ni, Sn and Si are in the interdendritic spaces. In the latter alloy, metallic Sn can be distinguished in the interdendritic spaces. E-waste alloys 2 and 3 show a good distribution of all alloying elements throughout the microstructure.

The hardness measured at 1 kg load for the four designed e-waste alloys is shown in Figure 8.

## RESULTS

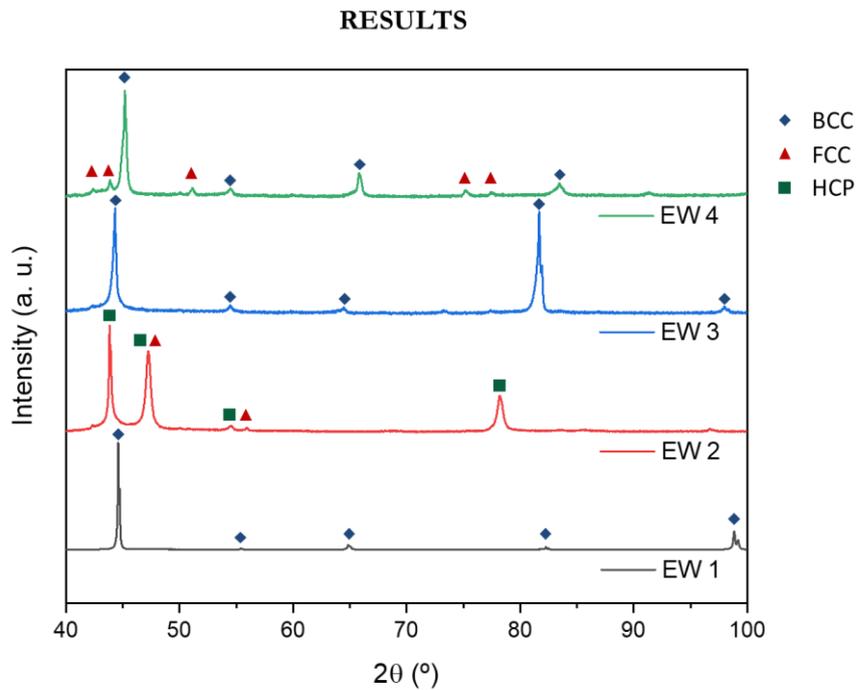

Figure 2. XRD diffraction patterns for the four studied e-waste alloys.

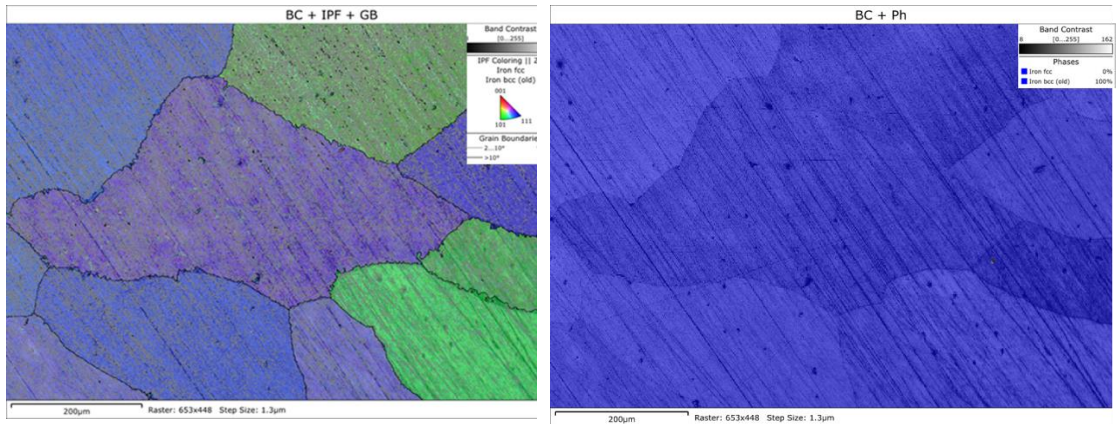
Figure 3. IPF-z map (left) and phases map (right) for the e-waste alloy 1.

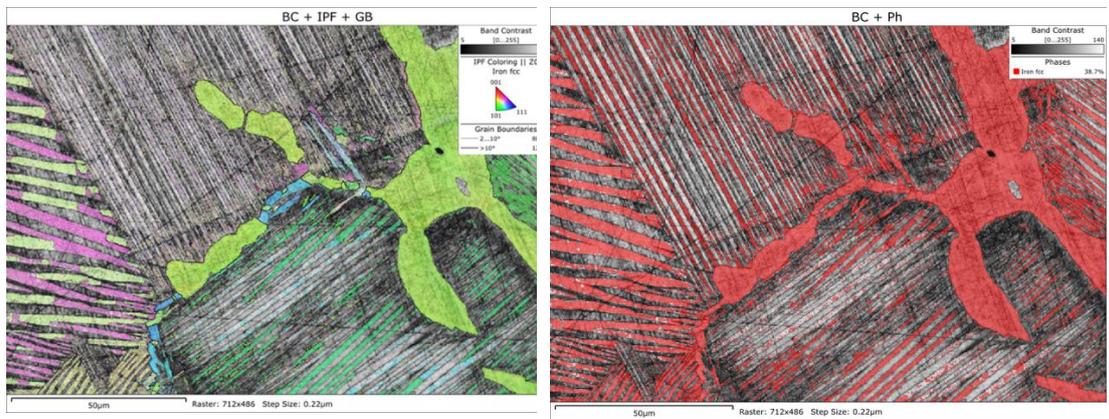
Figure 4. IPF-z map (left) and phases map (right) for the e-waste alloy 2.

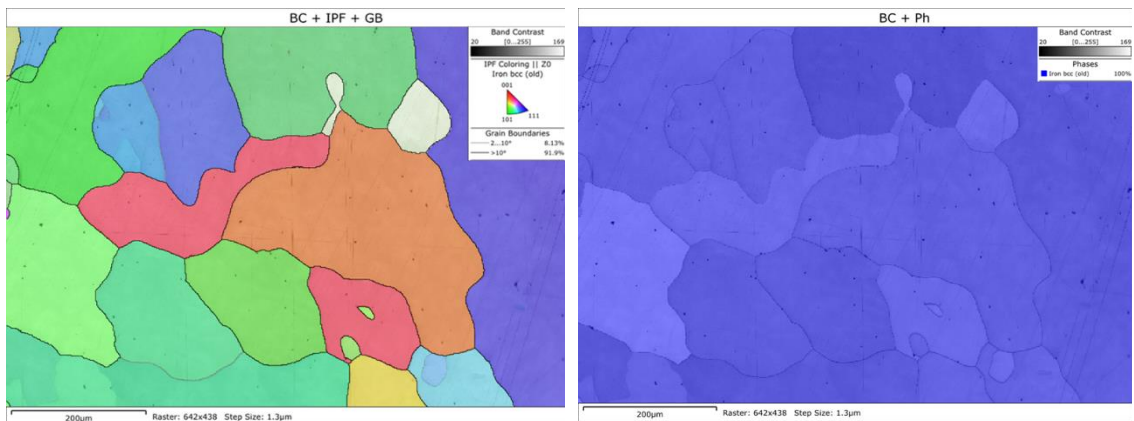
Figure 5. IPF-z map (left) and phases map (right) for the e-waste alloy 3.

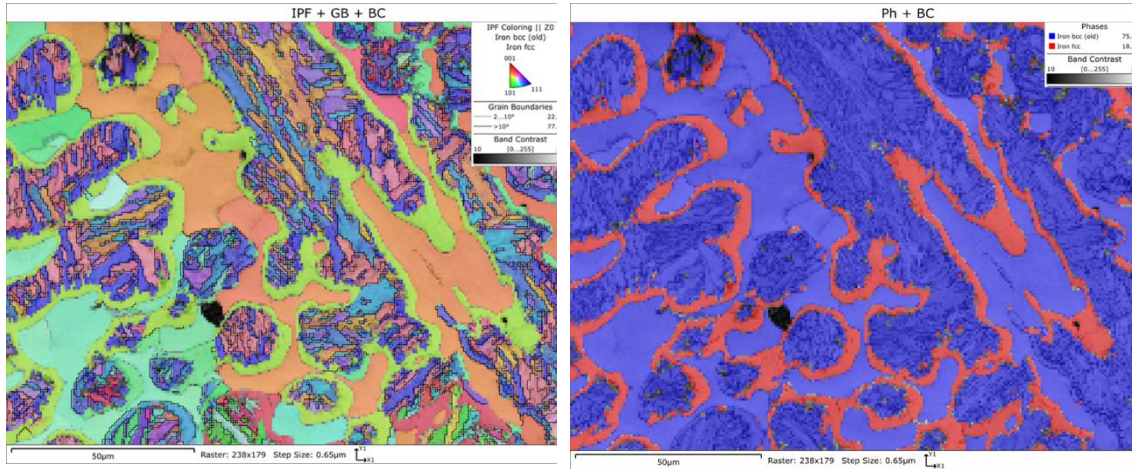

Figure 6. IPF-z map (left) and phases map (right) for the e-waste alloy 4.

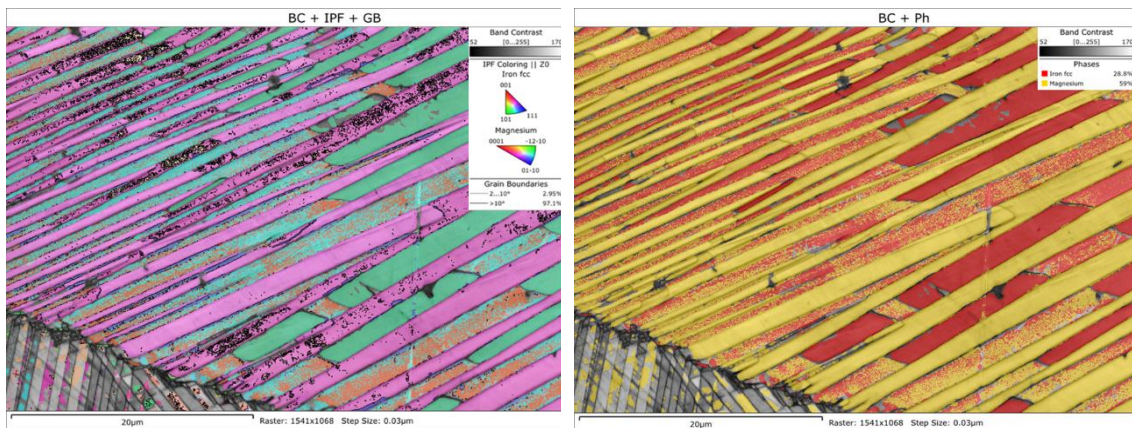

Figure 7. IPF-z map (left) and phases map (right) for e-waste alloy 2 at high magnification.

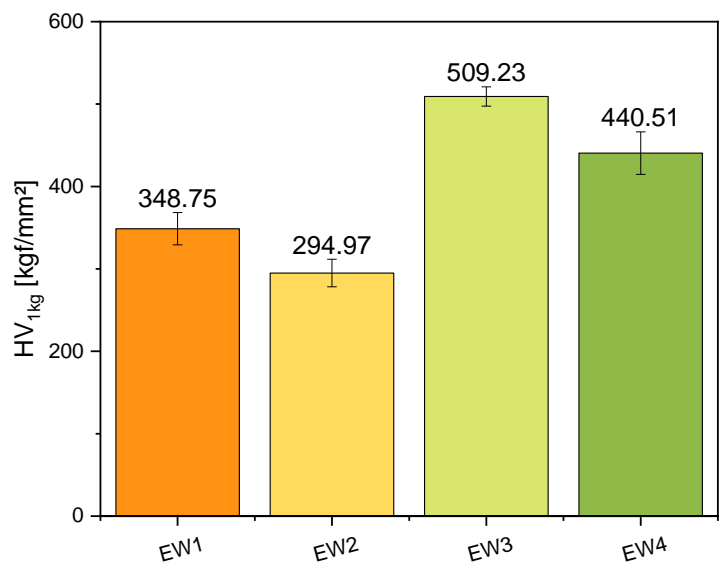

Figure 8. Hardness of the four e-waste alloys developed.

## 3. Discussion

This work aimed to demonstrate the feasibility of developing HEAs from the mixture of e-waste alloys obtained by direct leaching (or alternative extractive method). In this sense, and according to the prediction with different parameters established in the literature and compiled in the HEAPS software, we have identified four different candidate alloys using different sources of e-waste. We now discuss the agreement (or lack thereof) between predictions/models and alloys obtained by casting, as well as their resulting properties (i.e. hardness), in view of previous works in the literature.

### 3.1. Equilibrium calculations and phase predictions

For e-waste alloys 1, 3, and 4, CalPhaD calculation results (Figure 1) are in good agreement with experimentally assessed microstructure via EBSD and XRD, and in reasonable (yet partial) agreement with HEAPS phase predictions. For alloy 1, in addition to the predominant BCC phase, HEAPS also predicts the possible appearance of FCC phase, which also appears in CalPhaD calculations, however at low temperatures, while experiments show a monophasic BCC phase. In the case of alloy 3, HEAPS predicted the presence of a BCC phase and some possible intermetallic and TCP phases, in agreement with the CalPhaD predictions, which, however, also hint at the possible formation of sigma phase not predicted using HEAPS, and not identified in our experiments. For alloy 4, HEAPS predicted the presence of an FCC phase and some intermetallics. Meanwhile, CalPhaD calculations suggest the formation of a FCC phase first, followed by a possible transformation into a BCC phase (with complete transformation if full equilibrium is achieved), as well as the potential formation of several intermetallics. Experiments clearly show a (FCC+BCC) microstructure, hence in better agreement with CalPhaD (considering partial FCC to BCC transformation between 1000 and 1050°C).

Regarding alloy 2, both XRD and EBSD confirmed the presence of a HCP phase not predicted by either CalPhaD or HEAPS methods. Figures 4 and 7 show a typical off-eutectic microstructure, with a primary dendritic (FCC) phase and a secondary lamellar interdendritic eutectic structure. Besides the primary FCC phase, the second eutectic microconstituent is the identified HCP phase, while both CalPhaD and HEAPS suggest the presence of a BCC phase. In order to seek for a potential HCP phase, we calculated isopleth sections of the multicomponent phase diagram, changing the composition of Al – which may have a major influence on the eutectic microstructure[49] – at the expense of Ni. We performed those calculations considering all phases (Figure 9a), but also suspending the primary BCC-B2 phase (Figure 9b), in order to explore whether impeding its formation could potentially lead to the formation of a HCP phase. Figure 9a suggest that alloy 2 (29.2 at% Al) is hypereutectic (with respect to Al) with a primary BCC phase. Excluding the BCC phase from our calculations, Figure 9b places the alloy in the hypo-eutectic region, with a primary FCC phase, followed by the potential formation of a Heusler compound (with $L2_1$ structure), which is nearly equiatomic in Al and Co, with very minor solubility of Fe and Ni. Neither of these diagrams include a HCP phase, which suggests that the HCP phase is most likely not described in the CalPhaD database that we used (namely: TCHEA6). Moreover, we also explored the use of a Ni-based alloy database (TCNI8), which did not either hint at the potential formation of a HCP phase for alloy 2.

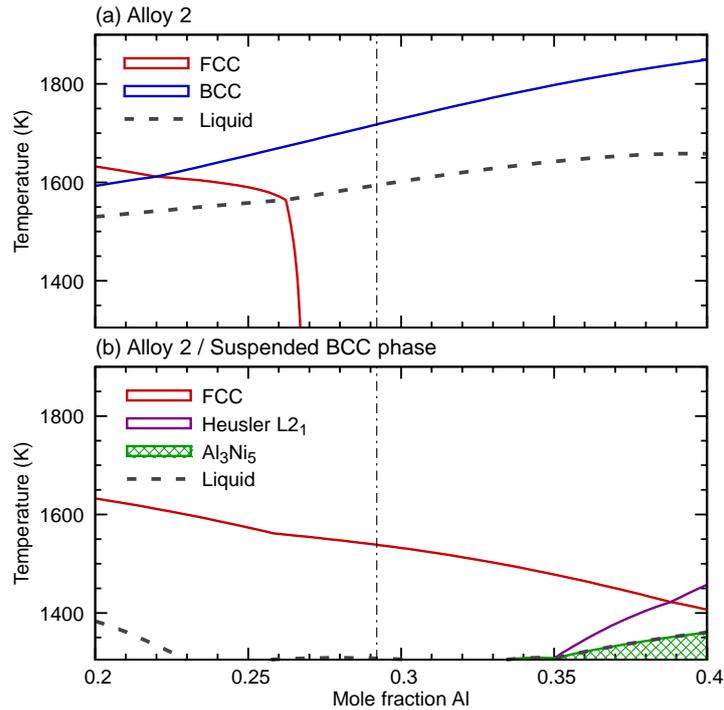

Figure 9. Isopleth phase diagram sections considering Al content at the expense of Ni: (a) For alloy 2 (as listed in Table 4), and (b) for alloy 2 with suspended BCC phase. The thin vertical dash-dotted line marks the nominal measured Al content of alloy 2.

### 3.2. Structural and microstructural features

In spite of the unexpected microstructure of alloy 2 discussed above, we can consider that the objective of this study has been achieved. All e-waste based HEAs exhibit a monophasic or biphasic microstructure without evidence of intermetallic phases. XRD and EBSD analyses are in complete agreement, thus providing confidence in the experimental phase identification. HEAPS-predicted phases only partially agree with the actual phases encountered in cast alloys. The formation of a single solid solution and the absence of Laves and Sigma phases were well predicted by HEAPS. As far as the microstructural features are concerned, with the exception of alloy 2, the alloys studied formed a dendritic microstructure, typical of cast alloys (particularly pronounced in alloy 4). The microstructure of e-waste alloy 2, which could not be predicted using either methods considered here (namely HEAPS and CalPhaD) is a typical off-eutectic HEA, where the dendrites correspond to the FCC phase and the eutectic microconstituent is formed by FCC and HCP.

e-waste alloys 1 and 3 have a monophasic BCC phase. Most monophasic BCC HEAs are composed of refractory metals and usually show good behaviour in terms of oxidation and corrosion with some ductility [50]. Here, we were capable of designing and synthesizing a full BCC structure without any refractory metals. The e-waste alloy 4 shows a mixed microstructure formed by FCC and BCC, obtained directly from casting without any further treatment. This alloy could, with a suitable manufacturing method and an optimised heat treatment, be competitive with other dual-phase HEAs typically obtained from Cantor alloy modifications. Dual FCC-BCC HEAs may exhibit special interfacial effects on plasticity and strengthening mechanisms [51] or good magnetic performance [52]. Finally, e-waste alloy 2 exhibits a typical off-eutectic microstructure, which might also be modified (i.e. enhanced) by heat treatments. Indeed, eutectic HEAs (EHEAs) are an emerging class of alloys with multiple promising applications, characterised by a high degree of microstructural customisation [53,54,55]. One of the most interesting aspect about the HEA developed from e-waste composition 2 is that its final microstructure is based on FCC and a HCP phase (potentially an intermetallic compound), whereas most EHEAs reported so far are composed of FCC and BCC (B2).

### 3.3. Hardness

Hardness is a useful property to assess the viability of alloys for structural applications, as a first approximation. It was reported in many HEAs of different compositions. Hardness levels measured in refractory HEAs can be higher than the values obtained here, from 500 HV [6,56] to 1000 HV [57]. The influence of some alloying elements such as Al or Mo on the microstructure of HEAs, producing eutectic alloys, has been widely studied, reaching values close to 600 HV for a high amount of Al [14] or 750 HV when the level of Mo is increased [58]. In [57] an interesting study is carried out on how hardness can be improved by substituting one of the Cantor alloy elements with different alloying elements. Therein, among 26 studied alloys, 12 have hardness values between 400 and 600 HV. In this context, and considering the hardness of some stainless steels, such as 316L (about 200 HV) and 17-4PH (about 400 HV), or even nickel-based superalloys, such as Hastelloy (250 HV) [14], the hardness values obtained in our e-waste alloys can be considered competitive – especially if we consider that the results were obtained directly from cast samples without any treatment to improve or refine their microstructure. Still, more comprehensive mechanical testing campaigns (e.g. tension, compression, fatigue, etc.) will be required in order to assess the applicability of the novel e-waste HEAs to perform in specific applications.

## 4. Summary and perspectives

Motivated by the exponential growth of e-waste products and by their unacceptably low recycling rate (< 20%), we have demonstrated the feasibility of designing competitive HEAs from mixing of typical e-waste multicomponent alloys and commodity alloys. We designed four multicomponent alloys produced from mixtures of e-waste alloy compositions typical, e.g., of smart phones, laptops, or Li-ion batteries (potentially mixed with some usual commodity alloys) that are candidates for high-performance HEAs.

The four candidate HEAs, made from e-waste alloy blends, were produced by vacuum arc-casting, characterised structurally, microstructurally, and mechanically. All four alloys can be considered as HEAs (solid solutions with one or two phases, without detrimental/brittle phases) with a real expectation of demonstrating a high-performance level.

These results allow us to affirm that it is perfectly feasible to obtain HEAs from multi-component alloy mixtures from e-waste. On the one hand, this proof-of-concept demonstration opens up new lines of research for the development of HEAs less reliant on critical/strategic elements, while also achieving properties that are difficult to reach with conventional alloys. On the other hand, it introduces a market for multi-component alloys from e-waste recycling, allowing a higher recyclability, as the extraction of multi-component alloys is cheaper and more efficient than the selective extraction of only a few metals.

In this paper, we have tried to demonstrate this hypothesis, but for e-waste recycling through the efficient manufacture of HEAs to be a viable and practical initiative, some challenges need to be addressed. The possible combination of thousands of potential starting alloys to achieve a given HEA with exceptional properties for a given application requires the development of machine learning tools that can, in a short time, decide on suitable mixing percentages. For these tools to have the ability to learn to enable convergent optimisations, major high-throughput manufacturing and characterisation campaigns will be required to focus on the desired properties for the different possible applications. Finally, the optimisation of leaching processes that allow the total extraction of elements from a given e-waste, as opposed to the current selective leaching, is also a challenge to be achieved.

From the point of view of both the development of new technologies and the need to optimise raw material resources, the need for new materials (e.g. HEAs) that push the limits of performance beyond the state of the art is imperative. Simultaneously, the need to achieve high levels of recycling of e-waste, which contains global reserves equivalent to those of nature, is imperative in a world where the use of raw materials has become a global issue. This work proposes a novel complementary way to address both of these urgent needs at once.

**Credit author statement**

José M. Torralba: conceptualisation, investigation, methodology, funding acquisition, supervision, writing – first draft; Diego Iriarte: investigation, data analysis, writing – review & editing; Damien Tourret: thermodynamic calculations, writing – review & editing ; Alberto Meza: investigation, methodology, writing – review & editing.

**Acknowledgements**

This work was supported by IMDEA Materials Institute's own funds. DT gratefully acknowledges support from the Spanish Ministry of Science through a Ramón y Cajal Fellowship (Ref. RYC2019-028233-I).

**Declaration of competing interest**

The authors declare that they have no known competing financial interests or personal relationships that could have appeared to influence the work reported in this article.

**Appendix A. Supplementary data**

The following is the supplementary data to this article: link to the file.